\documentclass[twocolumn,showpacs,preprintnumbers,prd,nofootinbib,floats,amssymb,floatfix]{revtex4}

\usepackage{amsmath}
\usepackage{amsfonts}
\usepackage{graphicx}
\usepackage{hyperref}

\begin{document}
\title{Inflation from IIB Superstrings with Fluxes}

\author{Erandy Ram\'irez}
\email{erandy@fis.cinvestav.mx}
\author{Tonatiuh Matos\footnote{IAC}}\email{tmatos@fis.cinvestav.mx}
\affiliation{Departamento de F{\'\i}sica, Centro de
Investigaci\'on y de Estudios Avanzados del IPN, A.P. 14-740,
07000 M\'exico D.F., M\'exico.}

\begin{abstract}
We study the conditions needed to have an early epoch of
inflationary expansion with a potential coming from IIB
superstring theory with fluxes involving two moduli fields. The
phenomenology of this potential is different from the usual hybrid
inflation scenario and we analyze the possibility that the system
of field equations undergo a period of inflation in three
different regimes with the dynamics modified by a Randall-Sundrum II
term in the Friedmann equation. We find that the system can
produce inflation and due to the modification of the dynamics, a
period of accelerated contraction can follow or preceed this
inflationary stage depending on the sign of one of the parameters
of the potential. We discuss on the viability of this model in a
cosmological context.
\end{abstract}

\date{\today}
\preprint{}
\maketitle

\section{Introduction}
The last 50 years have been one of the most fruitful ones in the
life of physics, the standard model of particles and the standard
model of cosmology (SMC) essentially developed in this period, are
now able to explain a great number of observations in laboratories
and cosmological observatories as never before. In only 50 years
great steps have been given in the understanding of the origin
and development of the universe. Nevertheless, many questions are
still open, for example the SMC contains two periods, inflation
and structure formation. For the understanding of the structure
formation epoch we need to postulate the existence of two kind of
substances, the dark matter and the dark energy. Without them, it
is impossible to explain the formation of galaxies and cluster of
galaxies, or the observed accelerated expansion of the universe.
On the other side, it has been postulated a period of inflation in
order to give an explanation for several observations as the
homogeneity of the universe, the close value of the density of the
universe to the critical density or the formation of the seeds
which formed the galaxies. However, there is not a theory that
unify this two periods, essentially they are disconnected from
each other.

In this work we study the possibility that superstring theory
could account for the unification between inflation and the
structure formation using a specific example. Recently, Frey
and Mazumdar \cite{Frey02}, were able to compactify the IIB superstrings
including 32 fluxes.
 In the context of the type IIB supergravity theory on the
${\bf T}^6/\mathbb{Z}_2$ orientifold with a self-dual three-form
fluxes, it has been shown that after compactifaying the effective
dilaton-axion potential is given by \cite{Frey02} %
\begin{eqnarray}
V_{dil}&=& \frac{M_P^4}{4(8\pi)^3}h^2\,e^{-2\Sigma_i\sigma_i}\,
\left[ e^{-\Phi^{(0)}}\cosh \left(\Phi-\Phi^{(0)}
  \right)  \right. \nonumber
  \\
  &+&\left.\frac{1}{2}\,e^{\Phi}(C-C^{(0)})^{2}-1\right],
  \label{eq:Poteff}
\end{eqnarray} where $h^2= {\frac{1}{6}} h_{mnp}h_{qrs} \delta^{mq} \delta^{nr}
\delta^{ps}$. Here $h_{mnp}$ are the NS-NS integral fluxes, the
superscript $(0)$ in the fields stands for the fields in the
vacuum configuration and finally $\sigma_i$ with $i=1,2,3$ are the
overall size of each factor ${\bf T}^2$ of the ${\bf
T}^6/\mathbb{Z}_2$ orientifold. This potential contains two main
scalar fields (moduli fields), the dilaton $\Phi$ and the axion
$C$. In \cite{mlgc}, the dilaton was interpreted as dark matter and
under certain conditions the model reproduces the observed
universe, $i.e.$, the structure formation. In this work we
investigate if the same theory could give an inflationary period
in order to obtain a unified picture between these two epochs. In
other words, in this paper we search if it is possible that the
same low energy Lagrangian of IIB superstrings with the scalar and
axion potential (\ref{eq:Poteff}) can give an acceptable
inflationary period. To deal with the fluxes, we work in the
brane representation of space-time, working with a RS-II
modification in the equations. Due to the presence of the
fluxes during this period, these models have the phenomenology of
the Randall-Sundum models \cite{rs1},\cite{rs},\cite{Frey02} and
we have chosen to work with the RS-II one, but the same type of
analysis could be done for the other model. The paper is organized
as follows. In section \ref{section:potential} we introduce an
appropriate parametrization of potential (\ref{eq:Poteff}) in order
to give unities to the physical quantities and to study the
cosmology of the low energy Lagrangian and give the field equation
to be solved. In section \ref{section:phenomenology} we solve the
equations in the different regimes and conditions. In section
\ref{num} we give the main results and in section
\ref{section:conclusions} we discuss some conclusions and
perspectives.\vfill

\section{The Potential}\label{section:potential}
\begin{figure}
\includegraphics[width=85mm, height=50mm]{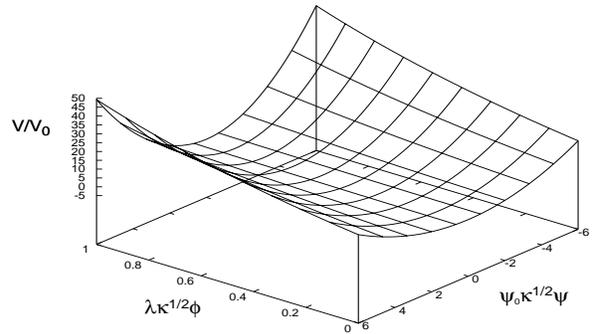}
\caption{Potential}\label{plotpot}
\end{figure}

In order to study the cosmology of this model, it is convenient to
define the following quantities
\begin{eqnarray}
\lambda\sqrt{\kappa}\phi&=&\Phi-\Phi^{(0)},\nonumber\\
V_0&=&\frac{M_P^4}{4(8\pi)^3}h^2\,e^{-2\Sigma_i\sigma_i}\,e^{-\Phi^{(0)}},\nonumber\\
C-C^{(0)}&=&\sqrt{\kappa}\psi,\nonumber\\
\psi_0&=&e^{\Phi^{(0)}},\nonumber\\
L&=&V_0\,(1-e^{\Phi^{(0)}}),
\end{eqnarray}
where $\lambda$ is the string coupling $\lambda =
e^{\langle\Phi\rangle}$, and $\sqrt{\kappa}=1/M_p$ 
with $M_p=({8\pi G})^{-1/2}$ 
the reduced Planck mass. With this new variables, the dilaton
potential transforms into
\begin{eqnarray} V_{dil}&=& {V_0}\, \left( \cosh \left( \lambda\sqrt{\kappa}\phi
  \right) -1\right)+\frac{1}{2}V_0\,{e^{\lambda\sqrt{\kappa}\phi }}{{\psi_0}}^{2}\kappa\psi^{2}+L\nonumber\\
&=&V+L.
 \label{pdil}
 \end{eqnarray}
where $L$ is interpreted as the cosmological constant, which we
will take as subdominant during the inflationary period. In what
follows we want to study the behaviour of this potential at early
times when the scalar field $\phi$ takes large values and study the
conditions that the parameters $\lambda$ and $V_0$ need to meet in
order to have inflation. Thus expressing the $\cosh$ function in
terms of exponentials and taking the limit $\phi$ big, we
arrive to the following expression for the potential:
\begin{equation}
V(\phi,\psi)=\frac{1}{2}V_0e^{\lambda\sqrt{\kappa}\phi}(1
+\kappa\psi_0^2\psi^2)-V_0,
\label{pinf}
\end{equation}
where we have taken into account that the cosmological constant is
much smaller than the coefficient of the potential, $L<<V_0$,
nevertheless, it remains a term $V_0$ acting during the
inflationary period like an extra ``cosmological constant" and
that we will address as a free parameter of the model. In contrast
with the usual hybrid inflation scenario \cite{l}, there is no
critical value for which this potential exhibits a phase
transition triggering the end of inflation (if any such process
occurs). The potential follows an exponential behaviour in the
$\phi$ field that prevents it from staying at a fixed value from
the start, i.e. it cannot relax at $\phi=0$ or at any other
different value (apart from infinity). A plot of the potential
illustrates this behaviour, see Fig. \ref{plotpot}.
We have checked that
there are no saddle points for this potential Eq.~(\ref{pinf}) and also
for potentials (\ref{eq:Poteff}) and (\ref{pdil}) 
\footnote{Many thanks to Anupam Mazumdar for this suggestion}, 
but a global minimum at $\psi=0$. 
Therefore, we will assume that there is some mechanism by which the
$\psi$ field rolls down to its minimum at $\psi=0$ oscillating
around it at the very early stages of evolution and that any
processes such as inflation took place afterwards. In this way, any
information concerning the evolution of $\psi$ is erased by the expansion
led by the $\phi$ field if inflation is to happen. Thus we can
work with the following expression for the potential:
\begin{equation}
V(\phi)=\frac{1}{2}V_0e^{\lambda\sqrt{\kappa}\phi}-V_0
\label{pphi}
\end{equation}
We will follow closely the analysis done by Copeland {\it et al.}
\cite {cllsw} and Mendes and Liddle \cite{ml} in order to obtain
the conditions for this potential to undergo inflation in the
cases when the scalar field or the $V_0$ term dominate the
dynamics as well as in the intermediate stage.
Our calculations are performed in the high-energy regime within
the slow-roll approximation since a potential slow-roll formalism
has already been provided for this scenario \cite{mwbh}.

\subsection{Field equations}

In the presence of branes, particularly in the RS-II scenario, the 
Friedmann equation changes from its usual expression to
\cite{rs}:
\begin{equation}
H^2=\frac{\kappa}{3}\rho\left(1+\frac{\rho}{\rho_0}\right),
\label{FE}
\end{equation}
where $H\equiv \dot{a}/a$, $a$ is the scale factor of the Universe, a dot
means derivative with respect to time  and
$\rho_0$ is the brane tension. The total density $\rho$ as well
as the equations of motion for the fields in the standard cosmology case
are deduced in \cite{mlul}:
\begin{equation}
\rho=\frac{1}{2}\dot{\phi}^2
+\frac{1}{2}\dot{\psi}^2e^{\lambda\sqrt{\kappa}\phi}
+V_{\phi}+V_{\psi}e^{\lambda\sqrt{\kappa}\phi}.
\label{rho}
\end{equation}
In our case,
\begin{equation}
V_{\phi}=\frac{1}{2}V_0e^{\lambda\sqrt{\kappa}\phi}-V_0,\hspace{0.5cm}
V_{\psi}=\frac{1}{2}V_0\kappa\psi_0^2\psi^2
\label{pots}
\end{equation}
which in the slow-roll approximation can be written as
\begin{equation}
H^2\simeq\frac{\kappa}{3}\left(V_{\phi}
+V_{\psi}e^{\lambda\sqrt{\kappa}\phi}\right)\left[1
+\frac{\left(V_{\phi}+V_{\psi}
e^{\lambda\sqrt{\kappa}\phi}\right)}{\rho_0}\right]
\label{FEsra}
\end{equation}

The equations of motion for both fields are given considering only the
presence of both scalar fields with no radiation
fluid, they are \cite{mlul}
\begin{eqnarray}
\ddot{\phi}+3H\dot{\phi}
+\frac{\partial{V_{\phi}}}{\partial{\phi}}
&=& \lambda\sqrt{\kappa}e^{\lambda\sqrt{\kappa}\phi}
\left(\frac{1}{2}\dot{\psi}^2-V_{\psi}\right) \nonumber \\
\ddot{\psi}+3H\dot{\psi}+\frac{\partial{V_{\psi}}}{\partial{\psi}}&=&
-\lambda\sqrt{\kappa}\dot{\phi}\dot{\psi}.
\label{ef}
\end{eqnarray}
Which in the slow-roll approximation are
\begin{eqnarray}
3H\dot{\phi}+\frac{\partial{V_{\phi}}}{\partial{\phi}}
&\simeq& -\lambda\sqrt{\kappa}e^{\lambda\sqrt{\kappa}\phi}V_{\psi} \nonumber \\
3H\dot{\psi}+\frac{\partial{V_{\psi}}}{\partial{\psi}}&\simeq& 0.
\label{efsra}
\end{eqnarray}
since the right-hand side of the second equation in (\ref{ef}) can be taken as
a kinetic term to the square.

There is an important consequence in the fact that we are considering the
$\psi$ field in the value corresponding to the minimum of its potential
at $\psi=0$. This leads to a system of equations that do not have source terms
and couplings of the fields and we end up with the same system
as for the usual RS-II modification. Otherwise, in order to
do a proper inflationary analysis, we would have to define new
expressions for the potential slow-roll parameters considering not only
the modification due to the RS-II term, but also to the couplings between the
$\psi$-potential and the $\phi$ in the Friedmann equation (\ref{FEsra})
and the source term in the first of equations (\ref{efsra}).

This procedure simplifies considerably the equations but avoids having a 
realistic analysis of the evolution of both fields including source terms 
and couplings, something that clearly must have an impact on the conditions 
to have inflation. Although a full analysis is undoubtedly required, we 
address the problem with these simplifications as a first insight before more
general assumptions are considered.

\section{Inflationary phenomenology of the model}\label{section:phenomenology}

The analysis that
follows will be confined to the high-energy limit of this model, which
simplifies calculations further. Consenquently, we have the following system
of equations within the slow-roll approximation and the high-energy limit:
\begin{eqnarray}
H\simeq \sqrt{\frac{\kappa}{3\rho_0}}V;\nonumber \\
3H\dot{\phi}+\frac{\partial{V_{\phi}}}{\partial{\phi}}\simeq 0
\label{srhefe}
\end{eqnarray}
which can be solved analitically. In this work we take the convention
$\dot{\phi}<0$, hence the $\phi$ field is a decreasing function of time.
The solutions of the field equations will be given in section \ref{sfe}.

The expressions for the potential slow-roll parameters for the RS-II cosmology
having the inflaton
field confined to the brane were
deduced by Maartens {\it et al.} \cite{mwbh}. For the high-energy limit
they are:
\begin{equation}
\epsilon\simeq \frac{1}{\kappa}\left(\frac{V'}{V}\right)^2\frac{\rho_0}{V};
\,\,\,\,\,\,\,\,\,
\eta\simeq\frac{1}{\kappa}\left(\frac{V''}{V}\right)\frac{\rho_0}{V}
\label{srp}
\end{equation}
where primes indicate derivatives with respect to the $\phi$
field. The slow-roll approximation is satisfied as long as the
slow-roll parameters defined previously accomplish the following
conditions :
\begin{equation}
\epsilon\ll 1,\,\,\,\,\,\,\,\,|\eta|\ll 1.
\label{src}
\end{equation}
The number of e-foldings of inflation in terms of the potential for this model
is given in our notation by \cite{mwbh}:
\begin{equation}
N\simeq -\frac{\kappa}{\rho_0}\int_{\phi_N}^{\phi_e}\frac{V^2}{V'}d\phi,
\label{N}
\end{equation}
where $\phi_N$ represents the value of $\phi$ N e-foldings of expansion
before the end of inflation and $\phi_e$ is the value of the field at the 
end of inflation.

For the potential (\ref{pphi}) we have
\begin{equation}
\epsilon=\frac{\lambda^2\rho_0}{4V_0}\frac{e^{2\lambda\sqrt
{\kappa}\phi}}{\left(\frac{e^{\lambda\sqrt{\kappa}\phi}}{2}-1\right)^3}
\label{cep}
\end{equation}
\begin{equation}
\eta=\frac{\lambda^2\rho_0}{2V_0}\frac{e^{\lambda\sqrt{\kappa}\phi}}
{\left(\frac{e^{\lambda\sqrt{\kappa}\phi}}{2}-1\right)^2}
\label{cet}
\end{equation}
\begin{eqnarray}
N\simeq \frac{2V_0}{\lambda^2\rho_0}\left[-\frac{1}{4}
\left(e^{\lambda\sqrt{\kappa}\phi_e}-e^{\lambda\sqrt{\kappa}\phi_N}\right)
\nonumber \right. \hspace{2cm} \\
\left.
+\left(e^{-\lambda\sqrt{\kappa}\phi_e}
-e^{-\lambda\sqrt{\kappa}\phi_N}\right)
+\lambda\sqrt{\kappa}(\phi_e-\phi_N)\right].
\label{cn}
\end{eqnarray}
The only way by which inflation can be ended in our model 
is by violation of the slow-roll
approximation with $\epsilon$ exceeding unity.

The value of $\phi$ at which $\epsilon$ becomes equal to unity
is
\begin{eqnarray}
\sqrt{\kappa}\phi_e=\frac{1}{\lambda}\ln\left\{\left[\frac{2\lambda^2\rho_0}
{3V_0}\right]
\left[\frac{4^{1/3}B^{1/3}}{2\lambda^2\rho_0}
\right.\right. \nonumber \hspace{2cm}\\
\left.\left.
+\frac{(6V_0
+\lambda^2\rho_0)4^{2/3}}{2B^{1/3}}
+\frac{(3V_0+\lambda^2\rho_0)}{\lambda^2\rho_0}\right]\right\}
\label{phiec}
\end{eqnarray}
with
\begin{eqnarray}
B=\lambda^2\rho_0\left[27V_0^2
+18V_0\lambda^2\rho_0+2\lambda^4\rho_0^2 \right. \nonumber \\
\left.
+(3V_0)^{3/2}\sqrt{4\lambda^2\rho_0+27V_0}\right].
\label{B}
\end{eqnarray}

We can rearrange the previous expression so that
\begin{eqnarray}
\sqrt{\kappa}\phi_e=\frac{1}{\lambda}\ln
\left\{\left[\frac{2\lambda^2\rho_0}{3V_0}\right]
\left[1+\frac{3V_0}{\lambda^2\rho_0}
+\frac{2^{2/3}B^{1/3}}{2\lambda^2\rho_0}
\right.\right. \nonumber \\
\left.\left.
+\frac{(6V_0+\lambda^2\rho_0)2^{1/3}}{B^{1/3}}\right]\right\}
\label{phief}
\end{eqnarray}
So we have that if
\begin{equation}
\left|\frac{3V_0}{\lambda^2\rho_0}+\frac{B^{1/3}}{2^{1/3}\lambda^2\rho_0}
+\frac{2^{1/3}(6V_0+\lambda^2\rho_0)}{B^{1/3}}\right|\ll 1,
\label{con}
\end{equation}
the $\phi$ field dominates the potential Eq.~(\ref{pphi}) and we will
have an exponential potential, which in the standard cosmology case
corresponds to power-law inflation, but not for the RS-II
modification.

The bound found before depends on the choice of values for $\lambda$ and
$V_0$. As we shall see in section \ref{fdr}, $\lambda$ is fixed once the
value of the brane tension is given. So actually Eq.~(\ref{con}) depends
only on the choice of  $V_0$. We use the computing packages Mathematica
and Maple to find numerically the values of $V_0$ ensuring the l.h.s. is 
real. This happens if:
\begin{eqnarray}
\lambda^2\rho_0<0, &\Longrightarrow& \hspace{0.5 cm}
V_0 \ge \frac{4}{27} \lambda^2\rho_0
\label{c1} \\
\lambda^2\rho_0>0, &\Longrightarrow& \hspace{0.5 cm}
V_0\le-\frac{1}{6}\lambda^2\rho_0
\label{c2}
\end{eqnarray}
Although both intervales guarantee a real value in the roots of condition 
(\ref{con}), this is not equivalent to have this condition satisfied in order
to have a field dominated region as we shall see later. 
From these expressions, the second case satisfies having a brane tension with
no incompatibilities with nucleosynthesis \cite{ml},\cite{cgs}, and we will
restrict further calculations to this posibility only.

\subsection{Density perturbations}

The field responsible for inflation produces
perturbations which can be of three types: scalar, vector
and tensor. Vector perturbations
decay in an expanding universe and tensor
perturbations do not lead to gravitational
instabilities producing structure formation.
The adiabatic scalar or density perturbations
can produce these type of instabilities through the
vacuum fluctuations of the field driving
the inflationary expansion. So they are
usually thought to be the seeds of the large scale structures
of the universe \cite{mfb}. One of the quantities that determines
the spectrum of the density perturbations is
$\delta_H$ which gives the density contrast at horizon
crossing (if evaluated at that scale). For the RS-II
modification and in our notation, this quantity is
\cite{ml},\cite{ll1}
\begin{equation}
\delta(k)_H^2\simeq\frac{\kappa^3}{75\pi^2}\frac{V^3}{V'^2}
\frac{V^3}{\rho_0^3}.
\label{dp}
\end{equation}
Evaluated at the scale $k=aH$.
The slow-roll approximation guarantees that $\delta_H$ is nearly
independent of scale when scales of cosmological interest are
crossing the horizon, satisfying the new COBE constrain updated to
WMAP3 \cite{wmap3} as
$\delta_H=1.9\times 10^{-5}$ \cite{lplm}. Here we take $60$
e-foldings before the end of inflation to find the scales of
cosmological interest.

We use equation (\ref{cn}),provided we know the value of $\phi_e$
given by equation (\ref{phiec}), to find the value $\phi_N$
corresponding to $N=60$, that is 60 e-foldings before the end of
inflation and evaluate $\delta_H$ as
\begin{equation}
\delta_H^2\simeq\frac{4\kappa^2}{75\pi^2}\frac{V_0^4}{\lambda^2\rho_0^3}
e^{-2\lambda\sqrt{\kappa}\phi_{60}}
\left(\frac{e^{\lambda\sqrt{\kappa}\phi_{60}}}{2}-1\right)^6
\label{dcc}
\end{equation}

The results are given in table \ref{t2} of section \ref{num}

\subsection{Field-dominated region}\label{fdr}

Considering the case when the $\psi$ field plays no role and
$\phi$ governs the dynamics of the expansion alone, from
Eq.~(\ref{pphi}), we have a potential of exponential type
resembling that of power-law inflation in the standard cosmology\cite{lm}. 
The first term of Eq.~(\ref{pphi}) dominates giving an
exponential expansion but not to power-law because
the dynamics of RS-II changes this condition. The
slow-roll parameters are given by:
\begin{equation}
\epsilon=\eta\simeq\frac{2\rho_0}{V_0}\frac{\lambda^2}
{e^{\lambda\sqrt{\kappa}\phi}}.
\label{srpp}
\end{equation}
In contrast with the standard cosmology where they are not only the same but
constant. The fact that the dynamics is
modified due to the Randall-Sundrum cosmology, allows the existence
of a value of $\phi$ that finishes inflation, since as we just saw,
the parameters show a dependence on $\phi$  and therefore, an evolution.

We have that in this regime, the value of $\phi_e$ corresponding to
the end of inflation is given by $\epsilon\simeq 1$. The $\simeq$ is used
because we are in the potential slow-roll approximation not the Hubble
one \cite{lpb}.

\begin{equation}
\sqrt{\kappa}\phi_{e}\simeq\frac{1}{\lambda}\ln
\left(\frac{2\lambda^2\rho_0}{V_0}\right)
\label{pep}
\end{equation}

Inserting this value in the expression for the number of
e-foldings (\ref{N}) for this regime and evaluating, we
obtain that $\phi_N$ with $N=60$ is
\begin{equation}
\sqrt{\kappa}\phi_{60}\simeq\frac{1}{\lambda}
\ln\left(122\frac{\lambda^2\rho_0}{V_0}\right),
\label{p60p}
\end{equation}
and we can evaluate the density contrast Eq~(\ref{dp}):
\begin{equation}
\delta_H^2\simeq\frac{61^4\kappa^2\lambda^6\rho_0}{75\pi^2}.
\label{dcp}
\end{equation}
One can obseve from this equation that given the value of
$\delta_H$ from observations, it
is possible to completely constrain $\lambda$ as
\begin{equation}
\lambda^6\simeq\frac{75\pi^2\delta_H^2}{61^4\kappa^2\rho_0}
\label{lambda}
\end{equation}
We have a dimensionles number that fixes one of the parameters
of the potential and can be contrasted with the value predicted
by this supergravity model when interpreted as Dark Matter \cite{mlgc}.

If we substitute the last equation into Eq.~(\ref{con}), we get in principle
a set of values for the constant $V_0$ that satisfy the field domination
condition.

\subsection{Vacuum energy-dominated regime}

We consider now the regime in which the second term of Eq.~(\ref{pphi})
dominates the dynamics. In this case the slow -roll parameters
Eqns.~(\ref{srp}) are:
\begin{equation}
\epsilon\simeq -\frac{\lambda^2\rho_0 e^{2\lambda\sqrt{\kappa}\phi}}{4V_0}
\label{epsv0}
\end{equation}
\begin{equation}
\eta\simeq\frac{\lambda^2\rho_0 e^{\lambda\sqrt{\kappa}\phi}}{2V_0}.
\label{etav0}
\end{equation}
We find ourselves here with the fact that $\epsilon$ is negative, that
is, with a period of deflation \cite{s}. Such behviour has already 
been predicted for these models before \cite{Frey02}.

It is necessary to point out that this is a consequence of the
modification of the dynamics. The definition of $\epsilon$ for the
Randall--Sundrum II cosmology in the high-energy limit is not positive
definite as in the standard cosmology case. Thus a stage of accelerated
contraction for this regime on the potential is only a result of the
modification in the field equations.

Following the evolution of the
dynamics with the potential Eq.~(\ref{pphi}), from a region where $\phi$
dominates, to a stage in which the energy $V_0$ drives the behaviour of
the expansion, we observe a primordial inflationary expansion that erases
all information concerning any process that the $\psi$ field might have
 undergone
under the influence of the potential Eq.~(\ref{pinf}). The intermediate
regime, in which both terms in Eq.~(\ref{pphi}) are of the same order, produces
further expansion. Finally the  field $\phi$ reaches a value on the
potential that comences a stage of accelerated contraction. This value
is obtained when the denominator in Eq.~(\ref{cep}) changes sign:
\begin{equation}
\sqrt{\kappa}\phi_d=\frac{\ln 2}{\lambda},
\label{phic}
\end{equation}
corresponding to the value of the vacuum-dominated regime.
Such process takes place when the field $\phi$ takes values below
$\sqrt{\kappa}\phi_d$. Eq.~(\ref{phic}) is equivalent to $V(\phi_d)=0$. Thus 
the balance of the terms
in Eq.~(\ref{pphi}) and the sign of $V_0$ determine the place where deflation
starts as the point where the potential crosses the $\phi$ axis.

In principle, there would not be a physical reason that
could prevent this deflationary stage to stop. But the argument
mentioned before, concerning the
modification of the dynamics applies again . We observe $\epsilon$ has
a dependence on the field and therefore
undergoes an evolution accordingly. The condition to end deflation is
$\epsilon=-1$, in oposition to inflation. In consequence, we can also find
from the first of equations (\ref{epsv0}) a value of $\phi$ corresponding to
this;
\begin{equation}
\epsilon=-1 \Rightarrow \,\,\,\,
\sqrt{\kappa}\phi_e=\frac{1}{2\lambda}\ln\left(\frac{4V_0}
{\lambda^2\rho_0}\right).
\label{eedf}
\end{equation}
Where this time, the subscript ``e'', indicates the end of deflation. This
value, as we can see, depends on $V_0$ and $\lambda$.

The choice of $V_0$ parameter will be  given in section \ref{num} where we give
different values  acording to the conditions we find in the following. We
shall see whether or
not an inflationary stage
takes place under the value of $\lambda$ found in the previous analysis.

\subsection{Intermediate Regime}
The intermediate regime corresponds to the region where both terms in
Eq~(\ref{pphi}) are of the same order. In order to obtain a bound on
the values of $V_0$ that satisfy the COBE constrain (\ref{dcc}), we need
to solve numerically Eqs.~(\ref{cn}) and (\ref{phiec}).Since we have that both
terms in Eq.~(\ref{pphi}) are important, this means that the exponential is
of ${\cal{O}}(1)$, thus we can expand it in Taylor series
as $\exp(\lambda\sqrt{\kappa}\phi)\simeq 1+\lambda\sqrt{\kappa}\phi$ and
arrive to a value of $\phi_e$ equal to
\begin{eqnarray}
\phi_e&=&\frac{(4 B)^{1/3}}{3\lambda V_0}
+\left(2+\frac{\lambda^2\rho_0}{3V_0}\right)
\frac{\lambda\rho_0 4^{2/3}}{\kappa B^{1/3}}
\nonumber \\ &&
+\frac{2\lambda\rho_0}{3\sqrt{\kappa}V_0}
+\frac{1}{\lambda\sqrt{\kappa}}
\label{phifi}
\end{eqnarray}
where $B$ now is given by
\begin{eqnarray}
B=\frac{\lambda^2\rho_0}{\kappa^{3/2}}
\left(18\lambda^2\rho_0V_0+27V_0^2+2\lambda^4\rho_0^2
\nonumber \right.\\ \left.
+\sqrt{(3V_0)^3(4\lambda^2\rho_0+27V_0)}\right)
\end{eqnarray}

In order to have a real scalar field, we find two bounds for
the values that $V_0$ can take on following from the
 roots in the previous expressions:
\begin{equation}
V_0<-\frac{1}{6}\lambda^2\rho_0, \hspace{1cm}
V_0>-\frac{4}{27}\lambda^2\rho_0\hspace{1cm} V_0>0.
\label{avp}
\end{equation}
The first bound coincides with the value given by Eq~(\ref{c2})
needed to have real values in condition (\ref{con}).
It is an upper bound for the allowed values
of $V_0$ in expression (\ref{phiec}). On the other hand,
we have from Eqns.~(\ref{phiec}) and (\ref{B}) that positive
values of $V_0$ also satisfy that there exists a real value of
$\phi_e$ in the intermediate regime. But $V_0$ cannot be 0 otherwise
many of the equations we have been looking at would be undefined. So we 
have an interval of allowed values of $V_0$ as:
\begin{equation}
V_0<-\frac{1}{6}\lambda^2\rho_0,\hspace{1cm} V_0>0.
\label{v1}
\end{equation}
If $V_0>0$, Eq.~(\ref{epsv0}) is negative and we have the period of deflation
mentioned before. But $V_0<-1/6\lambda^2\rho_0$ means that even in the
region of vacuum
domination $\epsilon$ can be positive and we return to the usual picture of
inflation. However, having chosen a value of
$V_0$ below this bound, Eqns.~(\ref{cep}) and (\ref{srpp}) become negative, 
thus giving a period of deflation translated to the epochs of field 
domination and the intermediate regime.

We can then choose values for $V_0$ below $-1/6\lambda^2\rho_0$ being
in a period of deflation for the first two regimes ending with a stage 
of inflation for the $V_0$ dominated region. 
Eqns.~(\ref{B}) and (\ref{phiec}) still have real values
because the approximation made in this section to find the upper limits of
$V_0$,
is a lower bound on Eqns.~(\ref{phiec}) and (\ref{B}). This is in fact 
redundant
since we have also said that both terms in the potential (\ref{pphi}) are
of the same order, so the expansion is valid for the general case.

Once the choice of $V_0$ is done, we
can solve numerically to find a value for $\phi_N$ in Eq.~(\ref{cn}), then it
is introduced into Eq.~(\ref{dp}) and can be accepted or rejected
depending on whether or not it fullfils the left-hand side. This value
depends on $\rho_0$, the brane tension, and we
present the results for different values of it considering that
in order to have no incompatibilities with nucleosynthesis the brane
tension must satisfy $\rho_0\ge 2 \mbox{MeV}^4$ \cite{ml}, the authors
take the number $1 \mbox{MeV}^4$, the difference arises due to the change
of notation. We also check that the choice on $\rho_0$ satisfies the
COBE constrain.

The results are
shown in table \ref{t2}.

\subsection{Vacuum-dominated region revisited}

Following the argument in the preceeding paragraph, it would be possible to
continue
with the usual analysis to find the value of $\phi$ that finishes inflation,
in the vacuum-dominated region
provided $V_0$ is negative and calculate the number of e-foldings and the
use Eq.~(\ref{dp}) in the region of vacuum domination to find a constrain
on $V_0$. We find from equation (\ref{epsv0}) that
\begin{equation}
\sqrt{\kappa}\phi_e=\frac{1}{2\lambda}
\ln\left(-\frac{4V_0}{\lambda^2\rho_0}\right),
\label{pev0}
\end{equation}
and from Eq~(\ref{N}) that
\begin{equation}
\sqrt{\kappa}\phi_{60}\simeq-\frac{1}{\lambda}
\ln\left[\left(-\frac{\lambda^2\rho_0}{4V_0}\right)^{1/2}
-\frac{30\lambda^2\rho_0}{2V_0}\right].
\label{p60v0}
\end{equation}
And finally from (\ref{dp}):
\begin{equation}
(-V_0)^{3/2}-60\lambda\rho_0^{1/2} V_0=
\frac{\sqrt{75}\pi\delta_H\rho_0}{\kappa}
\label{v0v0}
\end{equation}

This equation can be solved numerically to give a value of $V_0$
that is in accordance with the COBE constrain for the perturbations.
 As we shall see later, this process will not be applied to the
case of vacuum domination, since we find $\epsilon$ to be a decreasing
function of time, therefore for this case inflation never ends. Therefore 
it is not
possible to find a value of the parameter $V_0$ satisfying this condition
in the case of vacuum domination hence the values of $V_0$ coresponding to 
this regime are ruled out by 
observations. So in fact, the value of $\epsilon$
corresponding to 1 indicates in this case the place where inflation starts
to take place as from there onwards we will have that $\epsilon<1$

\subsection{Field Equations}\label{sfe}

In this subsection we solve the system of field equations. The numeric
results are presented in the next section. Integrating
Eqns.(\ref{srhefe}) for the
potential (\ref{pphi}) yields:
\begin{equation}
a(t)=a_0 \exp\left[-\frac{V_0}{\lambda^4 t\sqrt{3\kappa\rho_0^3}}
(\lambda^4\kappa\rho_0 t^2+24 e^{-2})\right]
\label{sft}
\end{equation}
and
\begin{equation}
\sqrt{\kappa}\phi(t)=-\frac{2}{\lambda}+\frac{1}{\lambda}
\ln\left(\frac{48}{\lambda^4\kappa\rho_0 t^2}\right).
\label{sfts}
\end{equation}
One can immediately see that the behaviour of the field does not
depend on the
value of the parameter $V_0$, but only on $\lambda$ whose value
is given by Eq.~(\ref{lambda}). From this solution and its plot, one can
check that indeed the field is a decreasing function of time, having
the same bahaviour regardless of the regime it is in. The scale factor
shows little dependance on the value of $V_0$ as shown in the plots. Following
an increasing behaviour for $V_0>0$. Figures \ref{sae-v0pos1},
\ref{sae-v0pos2}, \ref{sae-v0pos3} show that for
$V_0>0$, $\epsilon$ is a growing positive function which corresponds
to an inflationary stage as expected from the analysis of the previous
sections. The
acceleration factor is also shown, and one can observe that
for the range used in the plots $\ddot{a}/a$ changes sign before
$\epsilon$ reaches 1 in the same interval.

We have plotted for completeness the scale and acceleration factors 
as well as $\epsilon$ in Fig.\ref{sae-v0neg} for
a negative value of $V_0$. The scale and acceleration factors
change the sign of their slope at  $\lambda^2(\kappa\rho_0)^{1/2}t=1.8$
which is the
same value as that from Eq.~(\ref{phic}) indicating the start of the
vacuum-dominated regime. So one ends with a stage of inflation after
deflation in the other two regimes. We find that $\epsilon$ is
a positive decreasing function of time according with what was found. This 
means that although there is a
period of inflation, after deflation, it will never end and there is no
meaning in calculating the value of the potential from Eq.~(\ref{v0v0})
satisfying the COBE constrain  since there is no value of the field
corresponding to 60 e-folds before the
end of inflation. The case of vacuum domination with a negative potential
is not realistic for this model with our approximations.

The bounds that the field needs to meet in order to have
inflation for the intermediate regime are presented in table \ref{t4}. The
value of the parameter $V_0$ in the region of vacuum domination remains
unsconstrained, therefore it is not possible to give a bound on the
value of the field for the beginning of inflation.

\section{Results}\label{num}

Before starting with the numeric results for the three regions just
analized, we summarize what we have found so far. They are shown schematically
in table \ref{t1}.

\begin{table*}
\caption{Results for the sign of the first slow-roll parameter $\epsilon$
according to the choice
of $V_0$ for the three different regions of potential (\ref{pphi}).}
\begin{tabular}{|l|c|c|c|r|}\hline
Region    &  $V_0>0$   &   dynamics   &  $V_0<-\frac{1}{6}\lambda^2\rho_0$
&   dynamics \\
\hline
$\phi$-dominated    &  $\epsilon>0$, $\phi_e$ real  &   inflation &
$\epsilon<0$ & deflation \\
Intermediate &  $\epsilon>0$, $\phi_e$ real  &   inflation &  $\epsilon<0$
& deflation \\
Vacuum-dominated & $\epsilon<0$ & deflation &  $\epsilon>0$, $\phi_e$ real
&   inflation \\
\hline
\end{tabular}
\label{t1}
\end{table*}

Despite the fact that $V_0>0$ gives a positive value for Eq.~(\ref{srpp}),
it does
not correspond to the region of field domination. So one has to employ the
value of $V_0>0$ in Eqns. (\ref{cep}) and (\ref{phiec}) not in (\ref{pep}).
That is, in the intermediate regime. We have checked that
$V_0>0$ does not meet the condition (\ref{con}) even for very small values of
$V_0$ compared to unity. Instead, the smaller this parameter is, the closer
to 2 is Eq.~(\ref{con}). So we are left with only two regions where we
have inflation. The intermediate regime for $V_0>0$, and the region of
vacuum domination for $V_0<-1/6\lambda^2\rho_0^2$.

We solved numerically the corresponding equations in the intermediate
regime and found the values shown in table \ref{t2}. For this, we have taken
that $\kappa\simeq 25/m_{Pl}^2$ \cite{ml}.

\begin{table*}
\caption{Results for $V_0$ in the intermediate regime, for
three different values of $\rho_0$.}
\begin{tabular}{|c|c||c|c|c|r|}\hline
$\rho_0\times 10^6$, $(\mbox{eV}^4)$  &
$\lambda\times 10^{14}$ &
$V_0\times 10^{34}$, $\left({\mbox{eV}^4}\right)$ &
$\phi_e\times 10^{13}$, $\left(\mbox{eV}\right)$  &
$\phi_{60}\times 10^{13}$, $\left(\mbox{eV}\right)$  &
$\delta_H$, $\times 10^{-5}$ \\ \hline
  &
  &
$1.4$ &
$1.5$ &
$2.7$ &
$1.9$ \\
$2 $ &
$8.2$   &
$1.1$ &
$1.6$ &
$2.8$  &
$1.9$ \\
  &
  &
$0.9$ &
$1.6$ &
$2.8$  &
$1.9$ \\ \hline
  &
  &
$2.2$ &
$1.7$ &
$3.1$ &
$1.9$ \\
$4 $ &
$7.4$   &
$1.8$ &
$1.8$ &
$3.1$ &
$1.9$ \\
  &
  &
$1.4$ &
$1.9$ &
$3.2$ &
$1.9$ \\ \hline
  &
  &
$2.8$ &
$1.9$ &
$3.3$ &
$1.9$ \\
$6 $ &
$6.9$   &
$2.4$ &
$1.9$ &
$3.3$ &
$1.9$ \\
  &
  &
$1.9$ &
$1.9$ &
$3.4$ &
$1.9$ \\ \hline
\end{tabular}
\label{t2}
\end{table*}

Table \ref{t2} shows 3 values of the potential that are in good agreement with
the value of the density contrast $\delta_H$. The numbers that appear in the
third column conrrespond to  $1/100\lambda^2\rho_0$, $1/120\lambda^2\rho_0$
and $1/150\lambda^2\rho_0$ respectively. Bigger values in the denominators
seem to lead the decimals in the density contrast closer to
$1.9\times 10^{-5}$. We keep only these level of accuracy as a good 
approximation to the ideal value of the potential \cite{lplm}.

\begin{table*}
\caption{Bounds for the scalar field $\phi$ multiplied by the Planck mass,
for $V_0$ positive in eV units.}
\begin{tabular}{|c|c||c|c|}\hline
$\rho_0\times 10^6$, $(\mbox{eV}^4)$  &
$\lambda\times 10^{14}$ &
$V_0 $, $\left({\mbox{eV}^4}\right)$ &
$\phi > \times 10^{-15}\times m_{Pl}$, $\left(\mbox{eV}\right)$ \\ \hline
  &
  &
$\frac{1}{100}\lambda^2\rho_0$ &
$1.3$  \\
$2 $ &
$8.2$   &
$\frac{1}{120}\lambda^2\rho_0$ &
$1.3$ \\
  &
  &
$\frac{1}{150}\lambda^2\rho_0$ &
$1.4$ \\ \hline
  &
  &
$\frac{1}{100}\lambda^2\rho_0$ &
$1.4$ \\
$4 $ &
$7.4$   &
$\frac{1}{120}\lambda^2\rho_0$ &
$1.5$ \\
  &
  &
$\frac{1}{150}\lambda^2\rho_0$ &
$1.5$ \\ \hline
  &
  &
$\frac{1}{100}\lambda^2\rho_0$ &
$1.5$ \\
$6 $ &
$6.9$   &
$\frac{1}{120}\lambda^2\rho_0$ &
$1.6$ \\
  &
  &
$\frac{1}{150}\lambda^2\rho_0$ &
$1.7$  \\ \hline
\end{tabular}
\label{t4}
\end{table*}

\begin{figure}
\includegraphics[width=85mm, height=50mm]{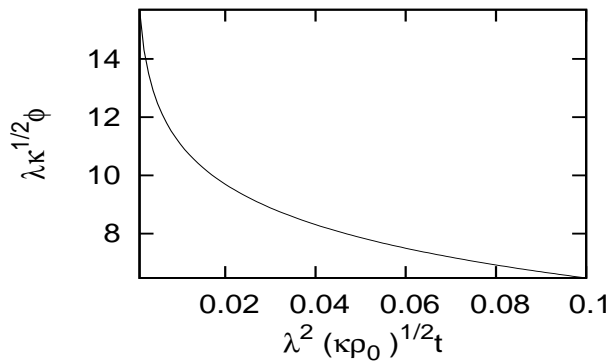}
\caption{The behaviour for the scalar field during
inflation.}\label{plotsf}
\end{figure}

\begin{widetext}

\begin{figure*}[htp]
\includegraphics[width=170mm, height=100mm]{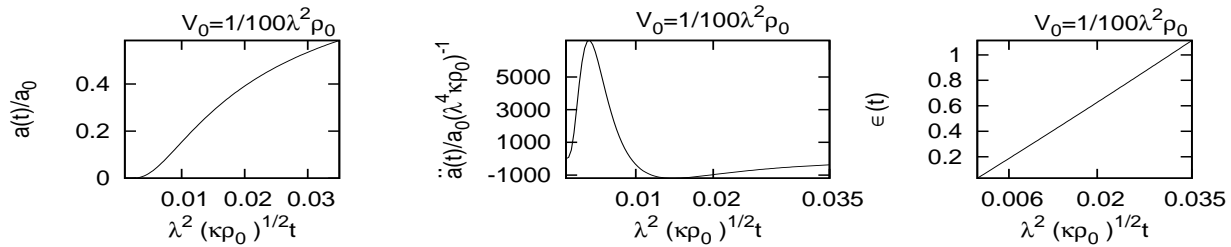}
\caption{On the left hand side the scale factor shows an
inflationary behaviour but the acceleration factor, on the center,
grows and decreases in the same interval. On the right hand side
we plot the inflationary parameter $\epsilon$. } \label{sae-v0pos1}
\end{figure*}

\begin{figure*}[htp]
\includegraphics[width=170mm, height=100mm]{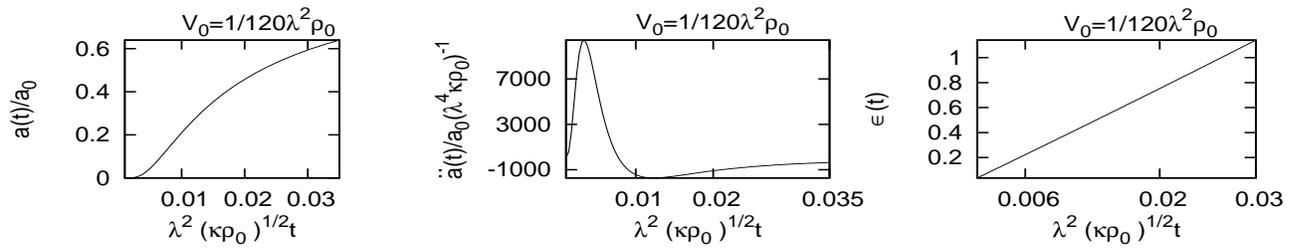}
\caption{In this case the scale factor (lhs) increases more
rapidly than the previous case and the acceleration factor
(center) reaches higher numbers within the same interval. On the
rhs, again we plot the $\epsilon$ parameter.} \label{sae-v0pos2}
\end{figure*}

\begin{figure*}[htp]
\includegraphics[width=170mm, height=100mm]{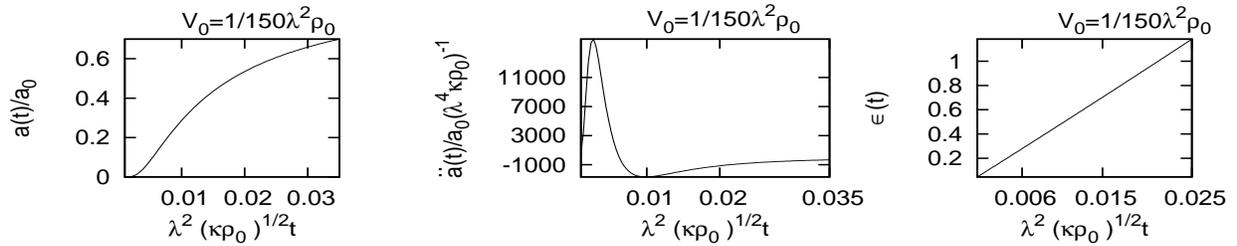}
\caption{Again in this case the scale factor (lhs) increases more
rapidly than in the first case and the acceleration factor
(center) reaches much higher numbers within the same interval. On
the rhs, again we plot the $\epsilon$ parameter.}
\label{sae-v0pos3}
\end{figure*}

\begin{figure*}
\includegraphics[width=50mm, height=50mm]{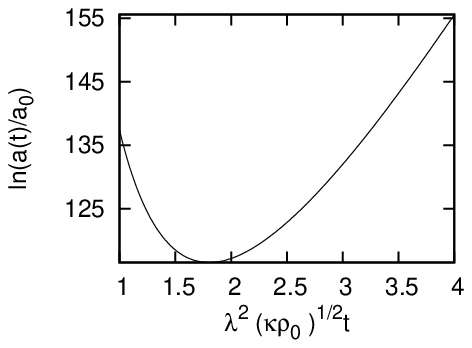}
\includegraphics[width=50mm, height=50mm]{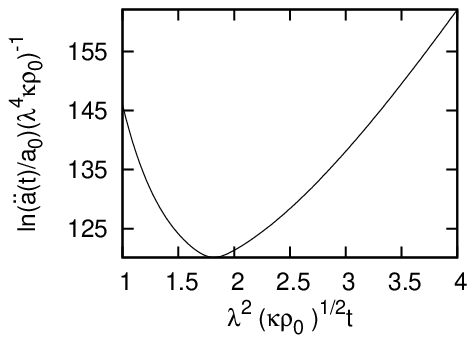}
\includegraphics[width=50mm, height=50mm]{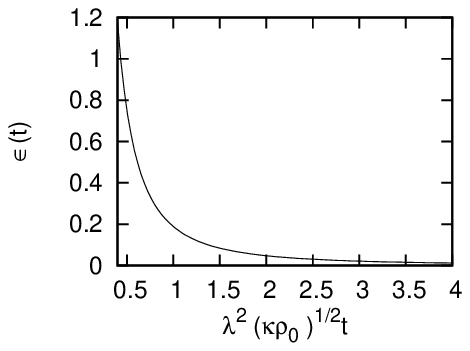}
\caption{The plot of the scale factor, the acceleration factor and
$\epsilon$ for $V_0$ negative, here we use
$V_0=-56\lambda^2\rho_0$ } 
\label{sae-v0neg}
\end{figure*}

\end{widetext}

\section{Conclusions}\label{section:conclusions}

In this work we have seen the conditions that the parameters of
potential (\ref{pphi}) have to fulfill in order to have
early universe inflation, using the same scalar field potential as
in \cite{mlgc} in order to have a unified picture between
inflation and structure formation. We find that the value of the
parameter $\lambda$ can be fixed in the region of field domination
and that there can be two possibilities for the sign of the
parameter $V_0$. Each of them determine different dynamics in the
evolution of the field equations. A positive sign leads to a
period of inflation followed by one of deflation, whereas the
opposite sign implies the contrary. In the first case, the values
of $V_0$ we have found as viable to meet the COBE constrain, are
not in agreement with those found in \cite{mlgc} by several orders
of magnitude. This behaviour seems to be generic in superstrings
theory, implying that if we would like to relate the moduli fields
with the inflaton, dark energy or dark matter, the model could fit
observations either during the inflationary epoch or during the
structure formation, but the challenge is to derive a model which
fit our observing universe during the whole history of the
universe. Otherwise, superstring theory have to give alternative
candidates for these fields and explain why we do not see the
moduli fields in our observations. Two important points to notice
are that the analysis employed in this work has been made with the
assumption that the field $\psi$ or axion does not play a
significant role in the dynamics and the slow-roll approximation
has simplified the equations further. This behaviour is observed
in the analysis carry out in \cite{mlgc}, where it is shown that
the axion field could remain as a subdominant field. Nevertheless,
even when this behaviour remains so until redshifts beyond
$10^{6}$, the axion could have a different behaviour beyond this
redshifts. This is indeed a very strong assumption which is not
well justified completely, but our work is a first approximation to
solve the problem and we are aware that a more general analysis
including the axion field needs to be done.

The second case corresponding
to vacuum domination has proved unrealistic to have a viable model
of inflation since this process does not end
and we do not have other mechanism to finish it as in the usual
hybrid inflation scenario. It would follow from this
that a more general analysis is needed in order to determine
whether the consequences of our assumptions are important or not.

\section{Acknowledments}

We are very thankful to Andrew Liddle and Anupam Mazumdar for helpful 
comments and revision of the paper. 
Many thanks to Abdel P\'erez-Lorenzana and Jorge Luis Cervantes-Cota for 
discussions. ER was supported by conacyt postdoctoral grant 54865.

\end{document}